\documentstyle[emulateapj]{article}

\def\ltsima{$\; \buildrel < \over \sim \;$}
\def\simlt{\lower.5ex\hbox{\ltsima}}
\def\gtsima{$\; \buildrel > \over \sim \;$}
\def\simgt{\lower.5ex\hbox{\gtsima}}
 
\received{}
\accepted{}
\journalid{}{}
\articleid{}{}
\slugcomment{To appear in ApJ-Letters}
\lefthead{}
\righthead{}

\begin{document}
   
\title{The BeppoSAX view of the hot cluster Abell 2319}
\author{S. Molendi\altaffilmark{1}, S. De Grandi\altaffilmark{2},
R. Fusco-Femiano\altaffilmark{3}, S. Colafrancesco\altaffilmark{4},
F. Fiore\altaffilmark{4,5}, R. Nesci\altaffilmark{6} and 
F. Tamburelli\altaffilmark{5}          
}

\altaffiltext{1}{Istituto di Fisica Cosmica, CNR, via Bassini 15,
I-20133 Milano, Italy}

\altaffiltext{2}{Osservatorio Astronomico di Brera, via Bianchi 46,
I-23807 Merate (LC), Italy}

\altaffiltext{3}{Istituto di Astrofisica Spaziale, CNR,
   via del Fosso del Cavaliere,
    I-00133 Roma, Italy}

\altaffiltext{4}{Osservatorio Astronomico di Roma Monteporzio Catone,
      Via Osservatorio 2, 
    I-00040 Monteporzio Catone (Roma), Italy}

\altaffiltext{5}{BeppoSAX Science Data Center, ASI,
    Via Corcolle, 19
    I-00131 Roma , Italy}

\altaffiltext{6}{Istituto Astronomico dell'Universit\`a di Roma 
La Sapienza, Via G.M. Lancisi 29, 
I-00161 Roma, Italy}

\begin{abstract}
We present results from a BeppoSAX observation of the rich cluster
Abell 2319.  The broad band spectrum (2-50 keV) of the cluster can be
adequately represented by an optically thin thermal emission model
with a temperature of 9.6$\pm$0.3 keV and a metal abundance of
$0.25\pm 0.03$ in solar units, and with no evidence of a hard X-ray
excess in the PDS spectrum.  From the upper limit to the hard tail
component we derive a lower limit of $\sim$0.04 $\mu$G for the
volume-averaged intracluster magnetic field.  By performing spatially
resolved spectroscopy in the medium energy band (2-10 keV), we find
that the projected radial temperature and metal abundance profiles are
constant out to a radius of 16$^{\prime}$ (1.4 Mpc).  A reduction of
the temperature of 1/3, when going from the cluster core out to
16$^{\prime}$, can be excluded in the present data at the 99$\%$
confidence level.  From the analysis of the temperature and abundance
maps we find evidence of a temperature enhancement and of an abundance
decrement in a region localized 6$^{\prime}$-8$^{\prime}$ NE of the
core, where a merger event may be taking place.  Finally, the
temperature map indicates that the subcluster located NW of the main
cluster may be somewhat cooler than the rest of the cluster.

\end{abstract}

\keywords{X-rays: galaxies --- Galaxies: clusters: individual 
          (Abell 2319)}

\section {Introduction}

Abell 2319 (hereafter A2319) is a rich cluster of galaxies located at
a redshift of z=0.056. Dynamical studies in the optical band (see
Oegerle et al. 1995, and references therein), have shown that the
cluster is actually composed of 2 main clumps. The main cluster and a
subcluster localized some 10$^{\prime}$ NW of the cD galaxy and behind
the main cluster.  At radio wavelengths A2319 is permeated by a radio
halo (Feretti, Giovannini \& B\"ohringer 1997, hereafter F97),
approximatively oriented in the NE-SW direction.  In the X-ray band a
temperature of $\sim $10 keV was measured with the Einstein MPC (David
et al. 1993).  Analysis of the ROSAT HRI image (F97, Peres et
al. 1998) shows that A2319 does not have the strongly peaked radial
profile typical of cooling flow clusters.  
A temperature map of A2319 has been obtained from ASCA data
(Markevitch 1996).  Recently, Irwin, Bregman \& Evrard (1999) have
used ROSAT PSPC data to search for temperature gradients for a sample
of galaxy clusters including A2319. While the ASCA measurements show a
radial temperature gradient (the temperature is about a factor of 2
smaller at about 2 Mpc than at the core), the ROSAT measurement is
consistent with a flat temperature profile.  

In this Letter we use BeppoSAX data to perform an independent
measurement of the temperature map of A2319.  We also present the
first abundance map of A2319 and the first measurement of the hard
(13-50 keV) X-ray spectrum of A2319.  The outline of the Letter is as
follows.  In section 2 we give some information on the BeppoSAX
observation of A2319 and on the data preparation.  In section 3 we
present the analysis of the broad band spectrum (2-50 keV) of A2319.
In section 4 we present spatially resolved measurements of the
temperature and metal abundance.  In section 5 we discuss our results
and compare them to previous findings.
Throughout this paper we assume H$_{o}$=50 km s$^{-1}$Mpc$^{-1}$ and
q$_{o}$=0.5.

\section {Observation and Data Preparation}
The cluster A2319 was observed by the BeppoSAX satellite (Boella et
al. 1997a) between the 16$^{th}$ and the 17$^{th}$ of May 1997.  We
discuss here data from two of the instruments onboard BeppoSAX: the
MECS and the PDS.  The MECS (Boella et al. 1997b, see Chiappetti et
al. 1998 for a particularly attractive interactive presentation) is
presently composed of two units, working in the 1--10 keV energy
range. At 6~keV, the energy resolution is $\sim 8\%$ and the angular
resolution is $\sim$0.7$^{\prime}$ (FWHM). The PDS instrument
(Frontera et al. 1997) is a passively collimated detector (about
1.5$\times$1.5 degrees f.o.v.)  working in the 13--200 keV energy
range.  Standard reduction procedures and screening criteria have been
adopted to produce linearized and equalized event files.  Both MECS
and PDS data preparation and linearization was performed using the
{\sc Saxdas} package under {\sc Ftools} environment.  The effective
exposure time of the observation was 3.8$\times$10$^4$ s (MECS) and
2.0$\times$10$^4$ s (PDS).  The observed countrate for A2319 was
0.908$\pm$0.006 cts/s for the 2 MECS units and 0.58$\pm$0.04 cts/s for
the PDS instrument.
All spectral fits have been performed using XSPEC Ver. 10.00.  Quoted
confidence intervals are 68$\%$ for 1 interesting parameter
(i.e. $\Delta \chi^2 =1$), unless otherwise stated.

\section{Broad Band Spectroscopy} 

We have extracted a MECS spectrum from a circular region of
16$^{\prime}$ radius (1.4 Mpc) centered on the emission peak. From the
ROSAT PSPC radial profile (see F97), we estimate that about 90$\%$ of
the total cluster emission falls within this radius.  The background
subtraction has been performed using spectra extracted from blank sky
event files in the same region of the detector as the source.  The PDS
background-subtracted spectrum has been produced by plain subtraction
of the ``off-'' from the ``on-source'' spectrum.  The spectra from the
two instruments have been fitted simultaneously, with an optically
thin thermal emission model (MEKAL code in the XSPEC package),
absorbed by a galactic line of sight equivalent hydrogen column
density, $N_H$, of 7.85$\times 10^{20}$ cm$^{-2}$.  A numerical
relative normalization factor among the two instruments has been added
to the spectral fit.  The reason is two-fold: a) the BeppoSAX
instrument response matrices employed in this Letter (September 1997
release) exhibit slight mismatches in the absolute flux calibration;
b) the PDS instrument field of view (1.3 degrees FWHM) covers the
entire emission from the cluster, while the MECS spectrum includes
emission out to 1.4 Mpc from the X-ray peak.  Taking into account the
mismatch in the absolute flux calibration, the vignetting of the PDS
instrument and the fraction of the emission falling outside of the
MECS extraction region, we estimate a normalization factor of 0.76. In
the fitting procedure we allow the normalization value to vary within
15$\%$ from the above value to account for the uncertainty in this
parameter.  The MEKAL model is found to fit the data adequately
($\chi^2 =$ 183 for 164 d.o.f.). The best fitting values for the
temperature and the metal abundance are respectively 9.6$\pm$0.3 keV
and 0.25$\pm$0.03, where the latter value is expressed in solar units.
The PDS data shows no evidence of a hard X-ray excess.  However, we
can derive a lower limit to the volume-averaged intracluster magnetic
field, B, responsible of the diffuse radio emission located in the
central region of the cluster. 
The radio halo spectrum shows an index $\alpha_r$ $\sim$0.92 in the
(408-610) MHz frequency range and $\sim$2.2 in the range (610-1400)
MHz; the radio flux is $\sim$1 Jy at 610 MHz (F97). From the PDS data
we can place a 90$\%$ confidence upper limits of 2.3$\times$10$^{-11}$
erg cm$^{-2}$ s$^{-1}$ and 2.0$\times$10$^{-11}$ erg cm$^{-2}$
s$^{-1}$ for a power-law spectrum with energy index 0.92 and 2.2,
respectively.  Relating the synchrotron radio halo flux to the X-ray
flux upper limits, assuming inverse Compton scattering of relativistic
electrons with the 3K background photons, we determine lower limits of
B of $\sim$0.04 $\mu$G and $\sim$0.035 $\mu$G, respectively.  The
equipartition magnetic field is estimated to be 0.48 $\mu$G (F97). We
determine also upper limits to the energy density of the emitting
electrons of $\sim$1.4$\times 10^{-12}$ erg cm$^{-3}$ and 9.3$\times
10^{-12}$ erg cm$^{-3}$ for $\alpha_r=0.92$ and 2.2, respectively,
using a size of $\sim$ 0.66 Mpc in radius for the radio halo.

\section{Spatially Resolved Spectral Analysis} 

A proper analysis of extended sources requires that the spectral
distortions introduced by the energy dependent PSF be correctly taken
into account.  
In the case of the MECS instrument onboard BeppoSAX the PSF, which is
the convolution of the telescope PSF with the detector PSF, is found
to vary only weakly with energy (D'Acri, De Grandi \& Molendi 1998).
This lack of a strong chromatic aberration results from the fact that
the telescope PSF degradation with increasing energy is
approximatively balanced by the improvement of the detector spatial
resolution.  Though we expect spectral distortions to be small, we
have taken them into account using the {\sc Effarea} program publicly
available within the latest {\sc Saxdas} release.  The {\sc Effarea}
program convolves the ROSAT PSPC surface brightness with an analytic
model of the MECS PSF to estimate the spectral distortions. A more
extensive description of the method may be found in D'Acri, De Grandi
\& Molendi (1998).  The {\sc Effarea} program also includes
corrections for the energy dependent telescope vignetting, which are
not discussed in D'Acri et al. (1998). The {\sc Effarea} program
produces effective area files, which can be used to fit spectra
accumulated from annuli or from sectors of annuli.

\subsection{Radial Profiles}

We have accumulated spectra from 6 concentric annular regions, with
inner and outer radii of 0$^{\prime}$-2$^{\prime}$,
2$^{\prime}$-4$^{\prime}$, 4$^{\prime}$-6$^{\prime}$,
6$^{\prime}$-8$^{\prime}$, 8$^{\prime}$-12$^{\prime}$ and
12$^{\prime}$-16$^{\prime}$.  The background subtraction has been
performed using spectra extracted from blank sky event files in the
same region of the detector as the source.  For the 5 innermost annuli
the energy range considered for spectral fitting was 2-10 keV, while
for the outermost annulus, due to the strong contribution of the
instrumental background in the 8-10 keV band, the fit was restricted
to the 2-8 keV range.
 
The ROSAT PSPC and HRI images of A2319 (F97) show excess emission,
with respect to a radially symmetric profile, in the NW and NE
direction.  Although the resolution of the MECS image (see figure 2)
is considerably poorer than that of the ROSAT images, evidence of the
excess is seen also in our data. The excess emission in the NW
direction is most likely associated to a subcluster identified from
velocity dispersion measurements (Oegerle et al. 1995), while the
structure in the NE direction is coincident with diffuse radio
emission observed at 20 cm (F97). To avoid possible contaminations to
the radially averaged spectra we have excluded data from the NW and NE
sectors of the third and fourth annuli. An analysis of the excluded
sectors is presented in the next subsection.  We have fitted each
spectrum with a MEKAL model absorbed by the galactic $N_H$ of
7.85$\times 10^{20}$ cm$^{-2}$.  In figure 1 we show the temperature
and abundance profiles obtained from the spectral fits.  The average
temperature and abundance for A2319 are found to be respectively:
9.7$\pm$0.3 keV and 0.30$\pm$0.03, solar units.  The temperature
profile is flat, with no indication of a temperature decline with
increasing radius.  A reduction of the temperature of 1/3, when going
from the cluster core out to 16$^{\prime}$ (1.4 Mpc), can be excluded
in the present data at the 99$\%$ confidence level.  The abundance
profile is consistent with being flat, although the large error
associated to the outermost annulus prevents us from excluding
variations in this region.  An abundance enhancement of a factor 2 in
the innermost region can be excluded in these data at more than the
99.99$\%$ level.
  
We have used the Fe K$_{\alpha}$ line as an independent estimator of
the ICM temperature.  We recall that for temperatures larger than a
few keV the Fe line comes mostly from the He-like Fe line at 6.7 keV,
and the H-like Fe line at 7.0 keV. As the temperature increases the
contribution from the He-like Fe line decreases while the contribution
from the H-like Fe line increases, thus the intensity ratio of the
He-like Fe line to the H-like Fe line can be used to estimate the
temperature. The MECS instrument does not have sufficient spectral
resolution to resolve the 2 lines, but it can be used to determine the
centroid of the observed line, which depends on the relative
contribution of the He-like and H-like lines to the observed line and
therefore on the gas temperature.  The position of the centroid of the
Fe K$_{\alpha}$ line is essentially unaffected by the spectral
distortion introduced by the energy dependent PSF and it depends only
weakly on the the adopted continuum model.  Thus it allows us to
derive an independent and robust estimate of the temperature profile.
Considering the limited number of counts available in the line we have
performed the analysis on 2 annuli with bounding radii,
0$^{\prime}$-8$^{\prime}$ and 8$^{\prime}$-16$^{\prime}$.  We have
fitted each spectrum with a bremsstrahlung model plus a line, both at
a redshift of z=0.056 (ZBREMSS and ZGAUSS models in XSPEC), absorbed
by the galactic $N_H$.  A systematic negative shift of 40 eV has been
included in the centroid energy to account for a slight
misscalibration of the energy pulseheight channel relationship near
the Fe line.  To convert the energy centroid into a temperature we
have derived an energy centroid vs. temperature relationship.  This
has been done by simulating thermal spectra, using the MEKAL model and
the MECS response matrix, and fitting them with the same model, which
has been used to fit the real data.  In figure 1 we have overlaid the
temperatures derived from the centroid analysis on those previously
obtained through the thermal continuum fitting. The two measurements
of the temperature profile are in agreement with each other.  The
moderate statistics available in the line does not allow us to place
very tight constrains on temperature gradients.
\subsection{Maps}

We have divided A2319 into 4 sectors: NW, including the subcluster
located behind A2319, SW, SE and NE. Each sector has been divided into
3 annuli with bounding radii, 2$^{\prime}$-4$^{\prime}$,
4$^{\prime}$-8$^{\prime}$ and 8$^{\prime}$-16$^{\prime}$. In figure 2
we show the MECS image with the sectors overlaid.  The background
subtraction has been performed using spectra extracted from blank sky
event files in the same region of the detector as the source.  We have
fitted each spectrum with a MEKAL model absorbed by the galactic
$N_H$.

In figure 3 we show the temperature profiles obtained from the
spectral fits for each of the 4 sectors.  In all the profiles we have
included the temperature obtained for the central region with radius
2$^{\prime}$.  Fitting each radial profile with a constant we derive
the following average sector temperatures: 9.8$\pm$0.4 keV for the NW
sector, 9.7$\pm$0.5 keV for the SW sector, 9.1$\pm$0.5 keV for the SE
sector and 10.1$\pm$0.5 keV for the NE sector.  All sector averaged
temperatures are consistent with the average temperature for A2319
derived in the previous subsection.  The $\chi^2$ values derived from
the fits indicate that all temperature profiles are consistent with
being flat. 
The temperature of the third annulus (bounding radii
4$^{\prime}$-8$^{\prime}$) of the NE sector is found to be higher than
the average cluster temperature at a significance level of $\sim
95\%$.  No significant temperature decrement is found in the third
annulus (bounding radii 4$^{\prime}$-8$^{\prime}$) of the NW sector,
where the subcluster is localized. Since this region includes the
emission of the subcluster as well as a fare amount of emission by
A2319, we have accumulated a spectrum from a smaller sector (see
figure 2), centered on the subcluster evidenced in figure 2. For this
region we find a temperature of 6.9$^{+1.2}_{-1.0}$ keV indicating
that the subcluster may have a somewhat lower temperature than A2319.

In figure 4 we show the abundance profiles for each of the 4 sectors.
In all profiles we have included the abundance obtained for the
central region with bounding radius 2$^{\prime}$.  Fitting each radial
profile with a constant we derive the following sector averaged
abundances: 0.29$\pm$0.04 for the NW sector, 0.29$\pm$0.05 for the SW
sector, 0.30$\pm$0.04 for the SE sector and 0.27$\pm$0.04 for the NE
sector.  All sector averaged abundances are consistent with the
average abundance for A2319 derived in the previous subsection.  The
$\chi^2$ values derived from the fits indicate that all abundance
profiles are consistent with being constant.  The abundance of the
third annulus (bounding radii 4$^{\prime}$-8$^{\prime}$) of the NE
sector is found to be smaller than the average cluster abundance at a
significance level of $\sim 95\%$.  The abundance of the subcluster
estimated from the third annulus (bounding radii
4$^{\prime}$-8$^{\prime}$) of the NW sector 0.22$\pm$0.07, or from the
smaller sector, shown in figure 2, 0.28$\pm$0.11, is consistent with
the average abundance of A2319.
 
\section{Discussion}

Previous measurements of the temperature structure of A2319 have been
performed by Markevitch (1996), using ASCA data, and by Irwin et
al. (1999), using ROSAT PSPC data.  
The average temperature,
10.0$\pm$0.7 keV, and abundance 0.3$\pm$0.08, reported by Markevitch
are in good agreement with those presented in this work.  
The temperature profile presented by Markevitch shows that the cluster
is isothermal, kT $\sim$ 11 keV, within 10$^{\prime}$ (0.9 Mpc) from
the core (however a slightly cooler spot associated to the NW
structure is found in this region), the temperature in the annulus
with bounding radii 10$^{\prime}$-20$^{\prime}$ (0.9-1.8 Mpc) is
smaller kT $\sim$ 8 keV, finally in the outermost bin,
20$^{\prime}$-24$^{\prime}$ (1.8-2.1 Mpc), the temperature falls to
about 4 keV, however the author remarks that this last point should be
considered less reliable than the others.  The BeppoSAX data shows no
evidence of a temperature decline within 16$^{\prime}$ (1.4 Mpc) from
the cluster core.  A reduction of the temperature of 1/3, when going
from the cluster core out to 16$^{\prime}$, can be excluded at the
99$\%$ confidence level.  To compare the ASCA measurement with ours,
we have converted the 90$\%$ confidence errors reported in figure 2 of
Markevitch (1996) into 68$\%$ confidence errors by dividing them by
1.65 (this is the factor that Markevitch \& Vikhlinin 1997 apply to
the 68$\%$ confidence errors of Briel \& Henry 1994 to convert them to
90$\%$ errors).  When we compare the individual data points covering
the same radial range we find that the temperature differences are
always significant at less than the 90$\%$ level.  To compare the
profiles we have derived the best fitting linear relationship for the
ASCA profile and compared it with our data finding a $\chi^2$ of 21.9
for 6 d.o.f. (probability of 1.3$\times 10^{-3}$).  We have also
compared the best fitting linear relation from our profile to the ASCA
profile finding a $\chi^2$ 12.4 for 3 d.o.f (probability of 6.1$\times
10^{-3}$).  Therefore, it seems that the two profiles are incompatible
with one another.  Recently Irwin, Bregman \& Evrard (1999) have used
ROSAT PSPC hardness ratios to measure temperature gradients for a
sample of nearby galaxy clusters, which includes A2139. From their
analysis they conclude that A2319 is isothermal out to 18$^{\prime}$
from the cluster core.  From the analysis of the temperature map we
find an indication of a smaller than average temperature,
6.9$^{+1.2}_{-1.0}$ keV , at the position of the subcluster. A similar
result, was also found in Markevitch (1996).

The average abundance of A2319 has been previously measured using ASCA
data, by Markevitch (1996), who finds 0.3$\pm$0.08, by Fukazawa et
al. (1998), who find 0.17$\pm$0.03 for Fe and 0.46$\pm$0.54 for Si,
and by Allen \& Fabian (1998), who find 0.33$\pm$0.06.  Our
measurement, like Markevitch's, given the adopted spectral ranges
(E$>$2.0 keV and E$>$2.5 keV respectively) is essentially an Fe
abundance measurement.  Our results are in agreement with the
measurements by Markevitch (1996) and by Allen \& Fabian (1998) and in
disagreement with the one presented by Fukazawa et al. (1998).  In
this Letter we present, for the first time, abundance profiles and
maps for A2319. The radial abundance profile is flat.  A2319 seems to
conforms to the general rule that non-cooling flow clusters do not
present abundance enhancements in their core.  In the third annulus
(bounding radii 4$^{\prime}$-8$^{\prime}$) of the NE sector we find
evidence of a temperature increase and of an abundance decrease, with
respect to the average values, both significant at about the 95$\%$
level.  F97, from the analysis of ROSAT PSPC and HRI images, find
evidence of excess emission in this region.  The same authors, from
the analysis of the 20 cm radio map of A2319, find also evidence of
diffuse radio emission in this region.  They argue that the presence
of the X-ray and radio structure may be the result of an on going
merger event in the NE direction.  Our measurement of a temperature
increase in the region supports their conjecture.  Indeed simulations
show that clusters undergoing a merger event should experience a
temperature enhancement in the merger region (e.g., Schindler \&
M\"uller 1993). The detection of an abundance decrease in the same
region may indicate that the subcluster merging with A2319 is poor in
metals.

\acknowledgments
We acknowledge support from the BeppoSAX Science Data Center.
We thank G. Zamorani for a critical reading of the manuscript.


\clearpage


\begin{figure}
\plotone{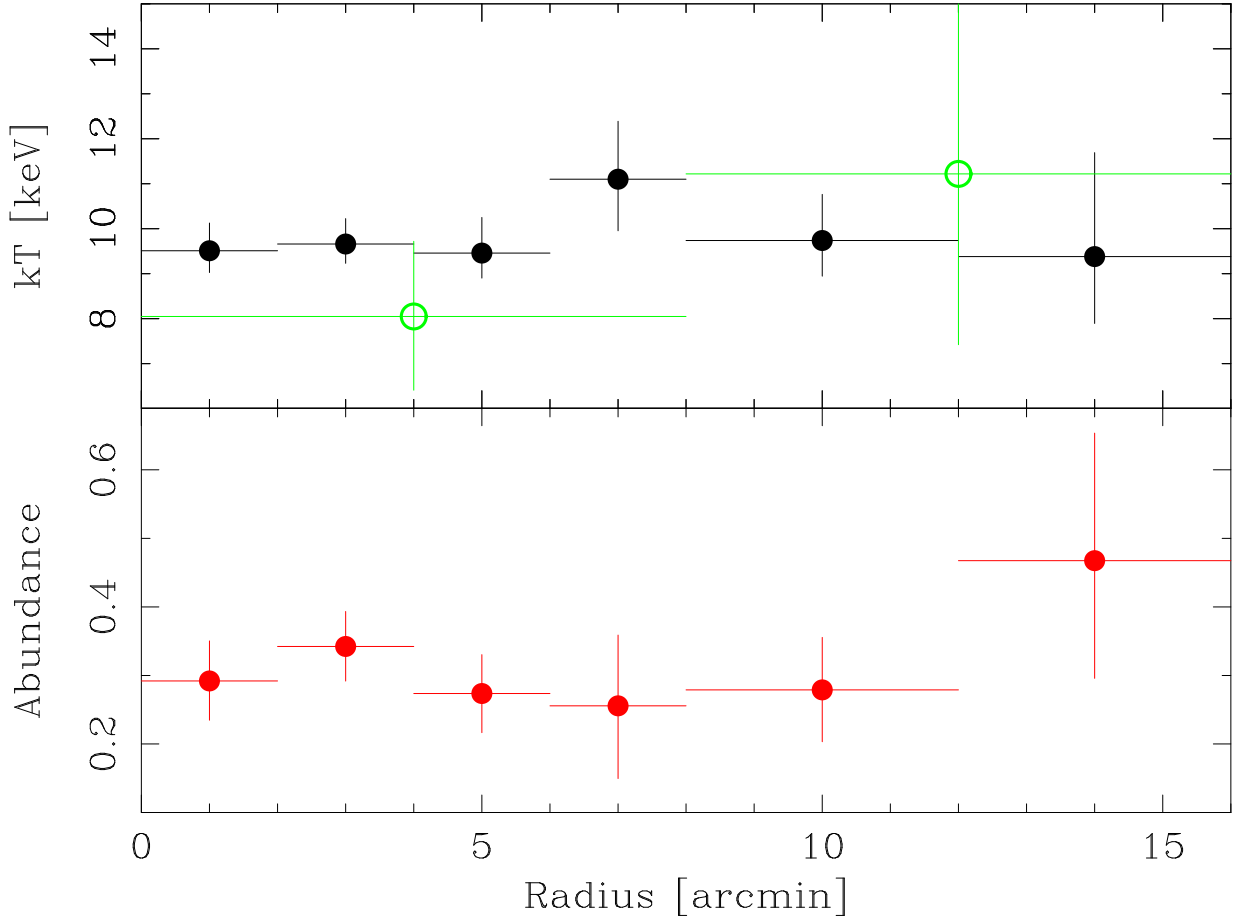}
\vskip -11cm
\caption 
{{\bf Top Panel}: projected radial temperature profile.  The filled
circles indicate temperatures obtained by fitting the continuum
emission. Open circles indicate temperatures estimated by the position
of the centroid of the Fe K$_{\alpha}$ line. 
{\bf Bottom Panel}: projected radial abundance profile.  
}
\end{figure}
\clearpage

\begin{figure}
\plotone{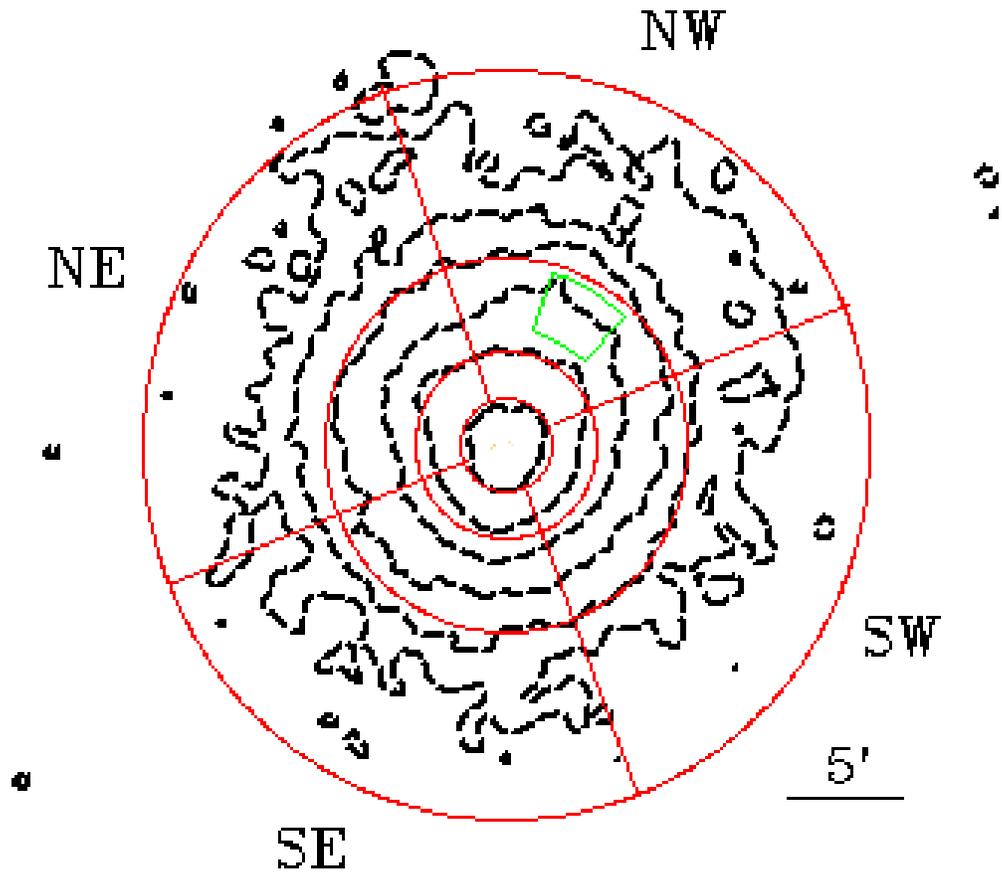}
\vskip -5cm
\caption
{BeppoSAX MECS image of A2319. Logarithmic contour levels are
indicated by the dashed lines. The solid lines show how the cluster
has been divided to obtain temperature and abundance maps.}
\end{figure}
\clearpage

\begin{figure}
\plotone{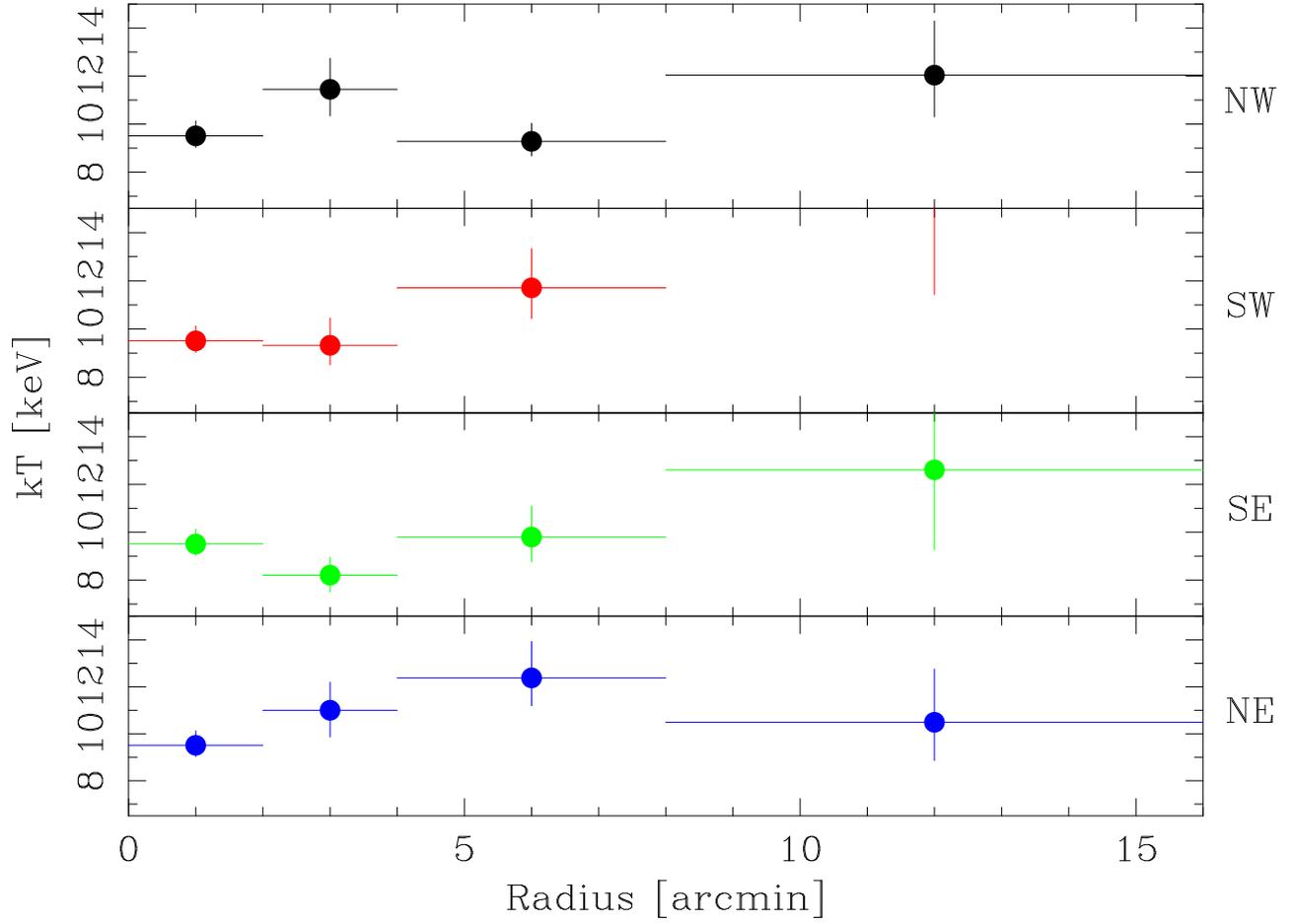}
\vskip -11cm
\caption 
{Radial temperature profiles for the NW sector (first panel), the SW
sector (second panel), the SE sector (third panel) and the NE sector
(forth panel).  The temperature for the leftmost bin is derived from
the entire circle, rather than from each sector. 
}
\end{figure}
\clearpage

\begin{figure}
\plotone{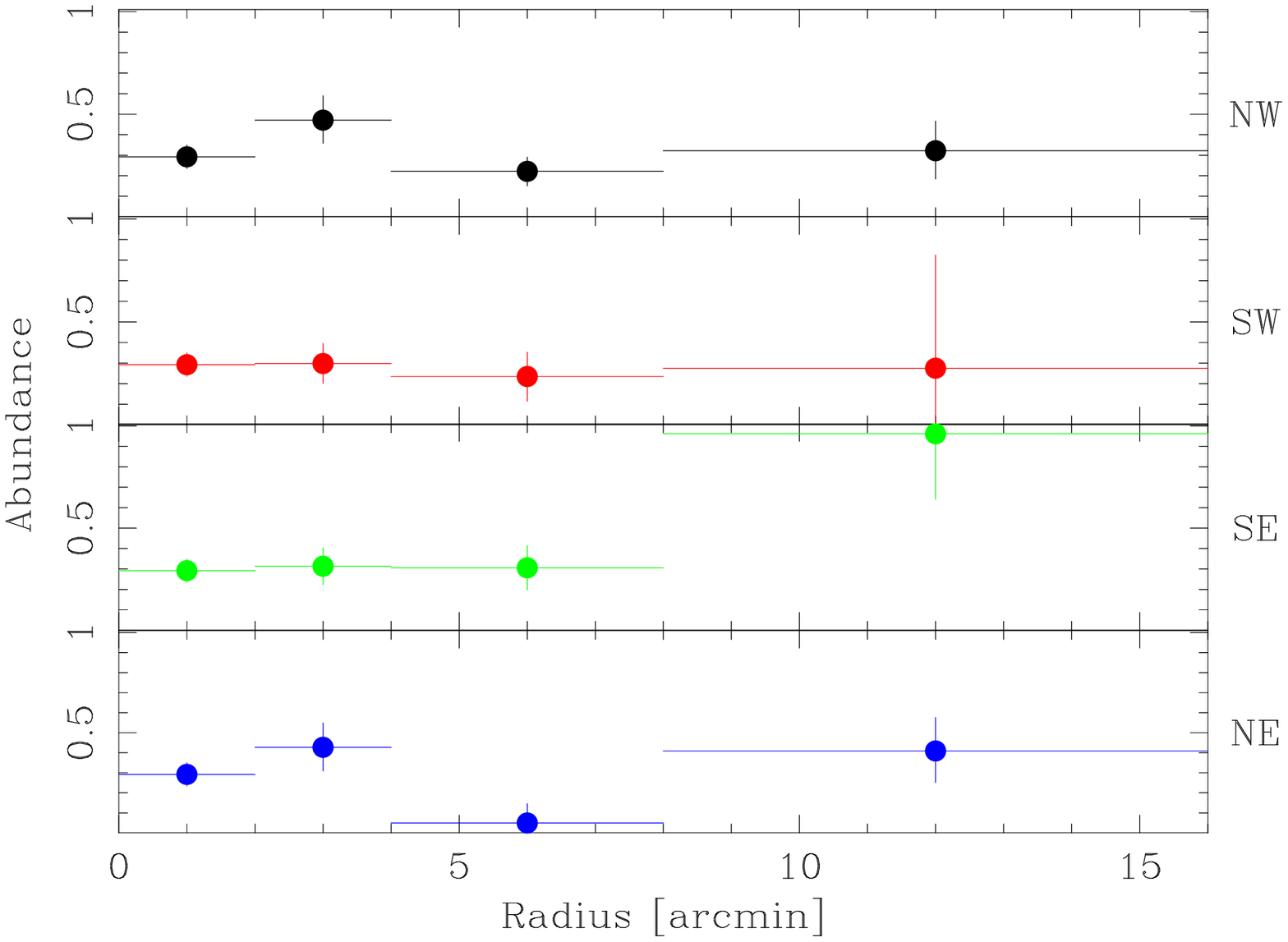}
\vskip -11cm
\caption 
{Radial abundance profiles for the NW sector (first panel), the SW
sector (second panel), the SE sector (third panel) and the NE sector
(forth panel).  The abundance for the leftmost bin is derived from the
entire circle, rather than from each sector. 
}
\end{figure}

\clearpage


\begin{thebibliography}{}


\bibitem{} Allen, S., W., \& Fabian, A., C. 1998, MNRAS, 297, L63
  
\bibitem{} Boella, G., Butler, R. C., Perola, G. C., Piro, L., Scarsi, L., 
\& Bleeker, J. A. M. 1997a, A\&AS, 122, 299 
\bibitem{} Boella, G., et al. 1997b, A\&AS, 122, 327 
\bibitem{} Briel, U. G., \& Henry, J. P. 1994, Nature, 372, 439 
\bibitem{} Chiappetti, L., et al. 1998
{\it http://sax.ifctr.mi.cnr.it/Sax/Mecs/tour.html}
\bibitem{} D'Acri, F., De Grandi, S., \& Molendi S. 1998, Nuclear Physics,
69/1-3, 581 (astro-ph/9802070)
\bibitem{} David, L. P., Slyz, A., Jones, C., Forman, W., Vrtilek, S. D., \& Arnaud, K. A. 1993, ApJ, 412, 479
\bibitem{} Feretti, L., Giovannini, G., \& B\"ohringer 1997, New Astronomy, 2, 501 (F97)
\bibitem{} Frontera, F., Costa, E., Dal Fiume, D., Feroci, M., Nicastro, L., 
 Orlandini, M., Palazzi, E., \& Zavattini G. 1997, A\&AS 122, 357
\bibitem{} Fukazawa, Y. et al. 1998, PASJ, 50, 187
\bibitem{} Fusco-Femiano, R., Dal Fiume, D., Feretti, L., Giovannini, G.,
 Grandi, P., Matt, G., Molendi, S., \& Santangelo, A. 1999 ApJL, 513, 21
\bibitem{} Irwin, J. A., Bregman J. N., \& Evrard A. E. 1999, ApJ in press, 
(astro-ph/9901406) 
\bibitem{} Markevitch, M. 1996, ApJL, 465, 1
\bibitem{} Markevitch, M., \& Vikhlinin, A. 1997, ApJ, 474, 84 
\bibitem{} Oegerle, W. R., Hill, J. M., \& Fitchett, M. J. 1995, AJ, 110, 32
\bibitem{} Peres, C. B., Fabian, A. C., Edge, A. C., Allen, S. W., 
Johnstone, R. M., \& White, D. A. 1998, MNRAS, 298, 416
\bibitem{} Schindler, S., \&  M\"uller, E. 1993, A\&A, 272, 137
\end{thebibliography}
\end{document}